\documentclass[]{aa}
\usepackage{psfig}
\usepackage{times}
\usepackage{graphics}

\def\la{\;
\raise0.3ex\hbox{$<$\kern-0.75em\raise-1.1ex\hbox{$\sim$}}\; }
\def\ga{\;
\raise0.3ex\hbox{$>$\kern-0.75em\raise-1.1ex\hbox{$\sim$}}\; }
 
\newcommand{\zabs}{$z_{\rm abs}\,$}
\newcommand{\zem}{$z_{\rm em}\,$}

\newcommand{\kms}{km~s$^{-1}\,$}
\newcommand{\cm}{cm$^{-2}\,$}
\newcommand{\cmm}{cm$^{-3}\,$}

\begin{document}

\title{Spectral shape of the UV ionizing background and \ion{O}{vi} absorbers at 
$z \sim 1.5$ towards \object{HS~0747+4259}\thanks{Based on 
observations obtained at the W. M. Keck Observatory and the HST}}
\author{
D.~Reimers\inst{1}
\and I.~I.~Agafonova\inst{2}
\and S.~A.~Levshakov\inst{2}  
\and H.-J.~Hagen\inst{1}
\and C.~Fechner\inst{1}
\and D.~Tytler\inst{3}  
\and D.~Kirkman\inst{3}  
\and\\ S.~Lopez\inst{4}
}
 
\institute{Hamburger Sternwarte, Universit\"at Hamburg, 
Gojenbergsweg 112, D-21029 Hamburg, Germany
\and Department of Theoretical Astrophysics, 
Ioffe Physico-Technical Institute, 194021 St. Petersburg, Russia
\and Center for Astrophysics and Space Science, University of
California, San Diego, MS 0424, La Jolla, CA 92093-0424, USA
\and Departamento de Astronomia, Universidad de Chile,
Casilla 36-D, Santiago, Chile
}

\offprints{S. A.~Levshakov, \protect \\lev@astro.ioffe.rssi.ru}
\date{received date; accepted date}

\abstract{}
{We report on high resolution spectra 
of the bright QSO \object{HS 0747+4259} (\zem = 1.90, $V = 15.8$) 
observed to search for intermediate redshift \ion{O}{vi}
absorption systems.}  
{The spectra were obtained by means of the Space Telescope Imaging
Spectrograph (STIS) at the Hubble Space Telescope (HST) and the
High Resolution Echelle Spectrometer (HIRES) at the W. M. Keck telescope.}
{We identify 16 \ion{O}{vi} systems in the range
  1.07 $\leq z \leq$ 1.87.  Among them, six systems with \zabs =
  1.46~--1.8 exhibit a sufficient number of lines of different
  ionic transitions to estimate the shape of the ionizing radiation
  field in the range 1~Ryd $< E < 10$~Ryd.  All recovered UV ionizing
  spectra are characterized by the enhanced intensity at $E > 3$ Ryd
  compared to the model spectrum of Haardt \& Madau (1996).  This is
  in line with the observational evidence of a deficiency of strong
  Ly$\alpha$ absorbers with $N$(\ion{H}{i}) $> 10^{15}$ \cm\, at $z <
  2$.  The UV background shows significant local variations: the
  spectral shape estimated at $z = 1.59$ differs from that obtained at
  $z = 1.81$ and 1.73.  A possible cause of these variations is the
  presence of a QSO/AGN at $z \simeq 1.54-1.59$ close to the line of
  sight.  No features favoring the input of stellar radiation to the
  ionizing background are detected, limiting the escape fraction of
  the galactic UV photons to $f_{\rm esc} < 0.05$.}
{}

\authorrunning{D. Reimers et al.}
\titlerunning{UV ionizing background and \ion{O}{vi} absorbers at $z
  \sim 1.5$} 
\maketitle

\section{Introduction}

The ionization state of the intergalactic medium (IGM) is maintained
by the metagalactic ionizing radiation field.  The current paradigm
considers the metagalactic UV field to be produced by the integrated
radiation of QSOs reprocessed by the clumpy IGM (Haardt \& Madau 1996,
hereafter HM96; Fardal et al. 1998).  Model calculations of the
spectral energy distribution (SED) of the UV background depend
crucially on the distribution of the Ly$\alpha$ forest lines~-- the
number of absorption lines per unit redshift and per unit \ion{H}{i}
column density.  This distribution is still poorly known.  For
instance, there is observational evidence suggesting a deficiency of
strong absorbers with $N$(\ion{H}{i}) $> 10^{15}$ \cm\ at $z < 2$
compared to higher redshifts (e.g. Kim et al. 2001).  These absorbers
are the main source of the \ion{He}{ii} continuum opacity and their
deficiency may result in a ionizing UV background much harder at
energies $E > 4$ Ryd than model predictions based on biased absorber
statistics (see HM96, Sect.5.14).  Another fact is that at $z < 2$ we
observe significant spatial variations in the absorption line number
density (Kim et al. 2002).

Closely related to the SED of the background ionizing radiation is the
problem of a possible galactic (stellar) contribution.  It is supposed
that the emissivity of QSOs decreases dramatically at $z < 2$, whereas
the input of galactic radiation to the ionizing background becomes
more significant (e.g., Bianchi et al. 2001; Bolton et al. 2005).
This conclusion is made from estimates of the mean intensity of the
ionizing background at 912 \AA.  However, a putative galactic
contribution would affect not only the intensity at the hydrogen
ionization threshold but also the shape of the ionizing spectrum
making it softer at $\lambda < 304$ \AA\, (Leitherer et al. 1999;
Giroux \& Shull 1997).  Thus, recovering the SED in the range $E > 4$
Ryd will enable us to estimate the relative contributions of QSOs and
galaxies to the ultraviolet background.

\begin{table*}[t]
\centering
\caption{Metal absorption-line systems toward  \object{HS 0747+4259}}
\label{tbl-t1}
\begin{tabular}{l@{ }ll}
\hline
\noalign{\smallskip}
$z_1$ & = 1.9157: & Ly$_{\alpha-\delta}$; 
              \ion{C}{iv}$_{1548,1550}$; \ion{O}{iv}$_{787}$  \\
$z_2$ & = 1.9070: & Ly$_{\alpha-\varepsilon}$; \ion{C}{iv}$_{1548,1550}$; 
                    \ion{O}{iv}$_{787}$ \\
$z_3$ & = 1.8766: & Ly$_{\alpha-\varepsilon}$; \ion{C}{iv}$_{1548,1550}$;
                    \ion{O}{iv}$_{787}$ \\
$z_4$ & = 1.8617: & Ly$_{\alpha-\gamma}$; \ion{C}{iv}$_{1548,1550}$;  
                  \ion{O}{iv}$_{787}$;  \ion{O}{vi}$_{1031}$  \\ 
$z_5$ & = 1.8533: & Ly$_{\alpha-\gamma}$; \ion{C}{iv}$_{1548,1550}$; 
                    \ion{O}{iv}$_{787}$; \ion{O}{vi}$_{1031}$ \\
$z_6$ & = 1.8073: &  Ly$_{1-12}$; \ion{C}{ii}$_{1334}$; 
                  \ion{Si}{ii}$_{1260}$; \ion{C}{iii}$_{977}$;
                  \ion{O}{iii}$_{832}$; \ion{Si}{iii}$_{1206}$;
               \ion{C}{iv}$_{1548,1550}$; \ion{O}{iv}$_{787}$;
                  \ion{Si}{iv}$_{1393,1402}$; \ion{N}{v}$_{1238}$;
                  \ion{O}{vi}$_{1031,1037}$\\
$z_7$ & = 1.7790: & Ly$_{\alpha,\beta}$; \ion{C}{iv}$_{1548,1550}$;  
                 \ion{O}{vi}$_{1031}$\\
$z_8$ & = 1.7744: &  Ly$_{\alpha-\varepsilon}$; \ion{C}{iii}$_{977}$;  
                  \ion{C}{iv}$_{1548,1550}$; \ion{O}{iv}$_{787}$;
                   \ion{O}{vi}$_{1031}$ \\
$z_9$ & = 1.7302: & Ly$_{1-8}$; \ion{C}{iii}$_{977}$; \ion{Si}{iii}$_{1206}$;  
                  \ion{O}{iii}$_{832}$;
               \ion{C}{iv}$_{1548,1550}$;  \ion{O}{iv}$_{787}$;
                  \ion{Si}{iv}$_{1393}$; \ion{O}{vi}$_{1031,1037}$ \\
$z_{10}$ & = 1.7154: & Ly$_{\alpha-\varepsilon}$;  
                       \ion{C}{iv}$_{1548}$; \ion{O}{vi}$_{1031}$ \\
$z_{11}$ & = 1.6331: & Ly$_{\alpha-\varepsilon}$; \ion{O}{iii}$_{832}$; 
                   \ion{C}{iv}$_{1548,1550}$; \ion{O}{vi}$_{1031}$ \\
$z_{12}$ & = 1.6134: & Ly$_{1-7}$; \ion{C}{iii}$_{977}$; 
                   \ion{O}{iii}$_{832}$; \ion{C}{iv}$_{1548,1550}$; 
                      \ion{O}{vi}$_{1031,1037}$  \\
$z_{13}$ & = 1.5955: & Ly$_{1-7}$;  \ion{C}{iii}$_{977}$; 
               \ion{O}{iii}$_{832}$; \ion{Si}{iii}$_{1206}$;
               \ion{C}{iv}$_{1548,1550}$; \ion{Si}{iv}$_{1393}$;
                \ion{O}{vi}$_{1031,1037}$  \\
$z_{14}$ & = 1.5401: & Ly$_{\beta-\gamma}$; \ion{C}{iii}$_{977}$;
               \ion{C}{iv}$_{1548,1550}$; \ion{Si}{iv}$_{1393,1402}$;
                \ion{O}{vi}$_{1031,1037}$  \\   
$z_{15}$ & = 1.4856: &  Ly$_{\beta}$; \ion{C}{iii}$_{977}$;
                \ion{C}{iv}$_{1548,1550}$; \ion{O}{vi}$_{1031}$  \\
$z_{16}$ & = 1.4648: & Ly$_{\beta-\varepsilon}$; \ion{C}{iii}$_{977}$;
                  \ion{C}{iv}$_{1548,1550}$; \ion{O}{vi}$_{1031,1037}$ \\
$z_{17}$ & = 1.2910 &  Ly$_{\alpha-\beta}$; \ion{C}{iii}$_{977}$;
                \ion{C}{iv}$_{1548,1550}$; \ion{O}{vi}$_{1031}$  \\
$z_{18}$ & = 1.1896 &  Ly$_{\alpha-\beta}$; \ion{C}{iii}$_{977}$;
                  \ion{C}{iv}$_{1548,1550}$; \ion{O}{vi}$_{1031}$  \\
$z_{19}$ & = 1.0778 &  Ly$_{\alpha}$; \ion{C}{ii}$_{1334}$;
                \ion{Si}{ii}$_{1260}$; \ion{Si}{iii}$_{1206}$;
                \ion{Al}{iii}$_{1854}$; \ion{C}{iv}$_{1548,1550}$; 
                \ion{Si}{iv}$_{1393,1402}$; \ion{O}{vi}$_{1031,1037}$\\
$z_{20}$ & = 1.0487 &  Ly$_{\alpha}$; \ion{C}{iv}$_{1548,1550}$ \\
$z_{21}$ & = 0.9277 &  Ly$_{\alpha}$; \ion{Si}{ii}$_{1260}$; 
                    \ion{Si}{iii}$_{1206}$; \ion{Al}{iii}$_{1854}$ \\ 
$z_{22}$ & = 0.2616 &  \ion{Mg}{ii}$_{2796,2803}$ \\
$z_{23}$ & = 0.2237 &  \ion{Mg}{ii}$_{2796,2803}$; \ion{Fe}{ii}$_{2600}$ \\
$z_{24}$ & = 0.2033 &  \ion{Mg}{ii}$_{2796,2803}$; \ion{Fe}{ii}$_{2382}$ \\
$z_{25}$ & = 0.0 &  \ion{Mg}{i}$_{2853}$; \ion{Mg}{ii}$_{2796,2803}$;
\ion{Ca}{ii}$_{3934,3969}$; \ion{Mn}{ii}$_{2594,2577}$;
\ion{Fe}{ii}$_{2600,2586,2382,2374,2344,2260}$ \\
\noalign{\smallskip}
\noalign{\smallskip}
\hline
\multicolumn{3}{l}{Note: \ion{O}{vi}~$\lambda\lambda1031,1037$ lines 
from systems with \zabs $> 1.8617$ fall in the gap between the HST/STIS and
Keck/HIRES spectra} 
\end{tabular}
\end{table*}

To probe the shape of the ionizing radiation, optically thin
absorption systems ($10^{13}$ \cm $\la N$(\ion{H}{i}) $< 10^{17}$ \cm)
with metal lines can be used (Chaffee et al. 1986; Bergeron \&
Stasinska 1986).  The ionization thresholds of ions usually observed
in the intervening absorbers (\ion{Si}{ii}-\ion{Si}{iv},
\ion{C}{ii}-\ion{C}{iv}, \ion{N}{iii}, \ion{N}{v}, \ion{O}{vi}) lie
between 1 and 10 Ryd and, thus, these boundaries limit the accessible
energy range.  Usually the shape of the ionizing continuum is
estimated by a trial-and-error method. Recently we developed a special
procedure based on techniques from the theory of experimental design
that allows us to restore the SED directly from the analysis of
absorbers with many metal lines (Reimers et al. 2005; Agafonova et
al. 2005).

In this paper we report on the SEDs obtained from the analysis of
metal absorbers with \zabs = 1.46-1.81 identified in the spectrum of
\object{HS 0747+4259}.  All absorbers studied reveal lines of
\ion{O}{vi}.  This ion can be produced either collisionally in high
temperature gas ($T_{\rm kin} > 10^5$~K) or by photoionization due to
the background ionizing radiation.  Particular interest in
collisionally ionized \ion{O}{vi} stems from the fact that \ion{O}{vi}
can trace the so-called `warm-hot intergalactic medium' (WHIM)
predicted in cosmological simulations (Cen \& Ostriker 1999; Dav\'e et
al. 2001).  Thus, the inferred shape of the ionizing continuum may
help us to understand which mechanism is preferable for the
interpretation of the observed \ion{O}{vi}.

We also estimate the frequency of \ion{O}{vi} absorbers in units of
absorption distance, $\Delta {\cal N}_{\rm OVI}/\Delta X$.

This paper is organized as follows.  The observations are described in
Sect.~2.   Sect.~3 contains a short description of the computational
methods used to analyze the absorption systems, which are studied in
detail in Sect.~4.  The results obtained are discussed and summarized
in Sect.~5.

\begin{figure*}[t]
\vspace{0.0cm}
\hspace{0.0cm}\psfig{figure=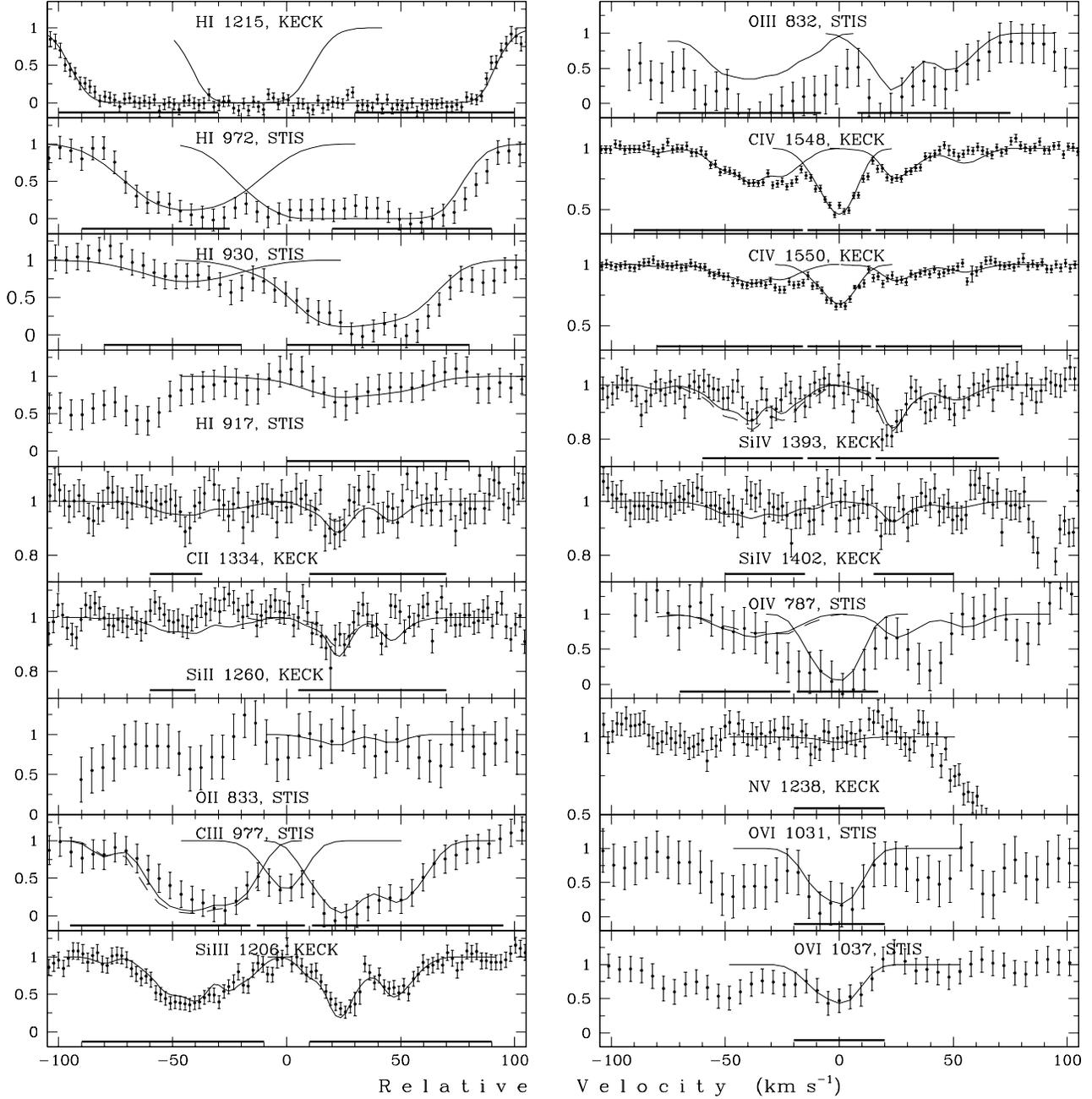,height=18.0cm,width=19.0cm}
\vspace{-0.7cm}
\caption[]{ Hydrogen and metal absorption lines associated with the
  \zabs = 1.8073 system toward \object{HS 0747+4259}.  Continuum
  normalized intensities are shown by dots, along with 1 $\sigma$
  error bars. The zero radial velocity is fixed at $z = 1.8073$.  The
  smooth curves are synthetic spectra calculated with the ionizing
  background S1 (Fig.~2).  The spectra are convolved with the
  corresponding spectrograph line-spread function. The physical
  parameters are listed in Table~2, Col.3, 5, and 6 (subsystems $A$,
  $B$, and $C$, respectively).  The dashed lines show synthetic
  spectra computed with the HM96 ionizing spectrum (Fig.~2), the
  physical parameters are given in Cols.2 and 4, Table~2.  Bold
  horizontal lines mark pixels included in the optimization procedure.
  The synthetic profiles of unmarked absorption lines were calculated
  in a second round using the velocity $v(x)$ and gas density $n_{\rm
    H}(x)$ distributions already obtained in the optimization
  procedure. }
\label{fig1}
\end{figure*}

\section{Observations}

The UV spectrum of the quasar \object{HS 0747+4259} (\zem = 1.90, $V =
15.8$; Reimers et al. 1998) was obtained with HST/STIS in September
2001 with a total exposure time of 54\,180 seconds over 19 orbits.
The spectrum was taken with the medium resolution NUV echelle (E230M)
and $0.2^{\prime\prime}\times0.2^{\prime\prime}$ aperture, providing a
resolution of $\lambda/\Delta\lambda \approx 30\,000$ (FWHM $\approx
10$ \kms).  The spectrum covers the range from 2136~\AA\ to 2970~\AA.
The data reduction was performed by the HST pipeline. After additional
inter-order background correction, the 12 individual exposures were
co-added. The resulting signal-to-noise S/N is 3-10 per pixel, lower
than expected because the quasar was fainter by a factor of 2 in the
UV compared to IUE observations (Reimers et al. 1998).

Another two integrations of \object{HS 0747+4259} were obtained with
HIRES on the Keck-I telescope: 3600 seconds on 2001 February 28, and
5400 seconds on 2001 March 01. The CCD was the single Tektronics
engineering grade with 2048 by 2048 24 micron pixels that was fixed in
HIRES from 1994 to mid-2004.  Both integrations used a single setup
covering 3060~\AA\ to approximately 4560~\AA\ with no gaps between the
39 spectral orders.  We used the C5 dekker which contains a 1.148
arcsec wide slit hat gives high throughput and approximately 8
\kms\ FWHM resolution.  Before each integration of the QSO, 300 second
integrations of the flux standard star G191 B2B were obtained.  A
Thorium-Argon lamp spectrum was used for the wavelength calibrations.
The spectrum was extracted and reduced using Tom Barlow's MAKEE
package, following procedures described in Kirkman et al. (2003) and
Suzuki et al. (2003).  The obtained S/N declines systematically into
the UV from $\sim 30$ at 4500~\AA\, to $\sim 5$ below 3200~\AA.

\begin{figure}[t]
\vspace{0.0cm}
\hspace{0.0cm}\psfig{figure=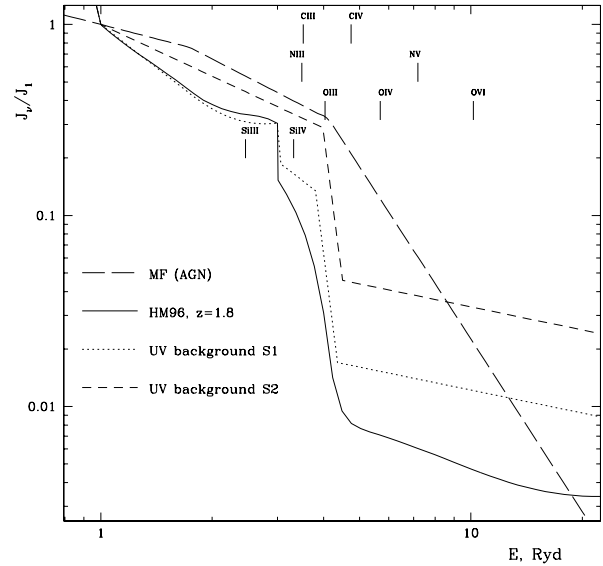,height=9.0cm,width=9.0cm}
\vspace{-1.2cm}
\caption[]{Schematic picture of the $z=2$ metagalactic (smooth line)
  and AGN-type (long-dashed line) ionizing backgrounds from Haardt \&
  Madau (1996) and Mathews \& Ferland (1987), respectively.  The
  spectra are normalized so that $J_\nu(h\nu =$ 1 Ryd) = 1.  The
  positions of ionization thresholds of different ions are indicated
  by tick marks. The results of the restoring procedure are shown by
  the dotted (S1) and short-dashed lines (S2). For details, see text. }
\label{fig2}
\end{figure}

\section{Computational methods}

The absorption systems are analyzed by means of the Monte Carlo
Inversion (MCI) procedure described in detail in Levshakov, Agafonova
\& Kegel (2000, hereafter LAK), and with modifications described in
Levshakov et al. (2002, 2003a,b).  Here we briefly outline the basics
needed to understand the results presented below in Sect.~4.

The MCI is based on the assumption that all lines observed in the
absorption system are formed in a continuous medium where the gas
density, $n_{\rm H}(x)$, and velocity, $v(x)$, fluctuate from point to
point (here $x$ is the space coordinate along the line of sight).

Further assumptions are that within the absorber the metal abundances
are constant, the gas is optically thin for the ionizing UV radiation,
and the gas is in thermal and ionization equilibrium.  This means that
the fractional ionizations of all ions are determined exclusively by
the gas density (or, equivalently, by the ionization parameter $U
\propto 1/n_{\rm H}$) and vary from point to point along the sightline.  
Since the ionization curves (i.e., the dependence of the ion
fraction on $U$) are different for different ions, the ionic line
profiles are different (non-similar) as well.  Another important fact
is that for a given point within the line profile the observed
intensity results from a mixture of different ionization states due to
irregular random shifts of the local absorption coefficient (see
Fig.~1 in LAK).

The fractional ionizations are determined by the SED of the background
ionizing radiation which is treated as an external parameter.

The radial velocity $v(x)$ and gas density $n_{\rm H}(x)$ are
considered as two continuous random functions which are represented by
their sampled values at equally spaced intervals $\Delta x$. The
computational procedure uses adaptive simulated annealing. The
fractional ionizations of all elements included in the analysis are
computed at every space coordinate $x$ with the photoionization code
CLOUDY (Ferland 1997).

In the MCI procedure the following physical parameters are directly
estimated: the mean ionization parameter $U_0$, the total hydrogen
column density $N_{\rm H}$, the line-of-sight velocity, $\sigma_{\rm
  v}$, and density, $\sigma_{\rm y}$, dispersions of the bulk material
[$y \equiv n_{\rm H}(x)/n_0$], and the chemical abundances $Z_{\rm a}$
of all elements involved in the analysis.  With these parameters we
can further calculate the column densities $N_{\rm a}$ for different
species, and the mean kinetic temperature $T_{\rm kin}$.  The mean gas
number density is related to the parameters of the gas cloud as [see
  eq.(28) in LAK]
\begin{equation}
n_0 = \frac{n_{\rm ph}}{U_0}(1 + \sigma^2_y)\; .
\label{E1}
\end{equation}
Here $n_{\rm ph}$ is the number density of photons with energies above
1 Ryd 
\begin{equation}
n_{\rm ph} = \frac{4\pi}{c\,h}\,J_{912}\,\int^\infty_{{\nu}_c}\,
\left( \frac{J_\nu}{J_{912}} \right)\,\frac{d \nu}{\nu}\; ,
\label{E2}
\end{equation}
where $c, h, \nu_c$, and $J_\nu$ are the speed of light, the Planck
constant, the frequency of the hydrogen Lyman edge, and the specific
intensity (in ergs cm$^{-2}$ s$^{-1}$ sr$^{-1}$ Hz$^{-1}$).  If the
mean gas number density $n_0$ is known then the line-of-sight
thickness $L$ of the absorber can be evaluated as $L = N_{\rm H}/n_0$.

An important issue in the analysis of the absorption systems is the
treatment of unidentified blends which can affect the line profiles
from the Ly$\alpha$ forest.  Since the method supposes that all ions
trace the same underlying gas density and velocity distributions, it
is possible to reconstruct both distributions using the unblended
parts of available lines.  To clarify which parts may be blended
several test runs with different arrangements of lines are carried out
until a self-consistent fit for all lines observed in the system is
found.

As mentioned above, the spectral shape of the ionizing radiation is
treated as an external parameter: some standard ionizing spectrum is
selected, corresponding ion fractions are calculated and then the MCI
analysis is carried out with these fractions inserted.  If the
selected spectrum fails to reproduce the observed line intensities or
some other physical inconsistencies arise (e.g. odd element abundance
ratios) the search for a more appropriate spectrum can be performed.
The corresponding computational technique is described in Reimers et
al. (2005) and Agafonova et al. (2005). It is based on the response
surface methodology from the theory of experimental design and
includes $(i)$ the parameterization of the spectral shape by means of
a set of variables (called `factors'), $(ii)$ the choice of a
quantitative measure (called the `response') to evaluate the fitness
of trial spectral shapes, and $(iii)$ the estimation of a direction
(in the factor space) which leads to a spectrum with better fitness.
The optimal (best fitness) spectral shape is that which allows us to
reproduce the observed intensities of all lines detected in the
absorption system without any physical inconsistencies.

How well the spectral shape can be recovered depends on the number of
metal lines involved in the analysis: the more lines of different
ionic transitions of different elements are detected in the absorption
system, the more constrained is the allowable set of shapes. When only
a few lines are detected, they can be used to estimate the shape in
some restricted energy interval. For example, lines of \ion{C}{iii},
\ion{C}{iv} and \ion{O}{vi} allow us  in some cases to estimate
the depth of the \ion{He}{ii} break ($E \geq 4$ Ryd).  Any other a
priori information concerning absorbers under study should be
considered as well in order to distinguish between possible solutions.

\begin{table*}[t]
\centering
\caption{ Physical parameters of the \zabs = 1.8073 metal absorber
  towards \object{HS 0747+4259} derived by the MCI procedure with the
  Haardt \& Madau and modified UV background spectra (marked,
  respectively, by HM96 and S1 in Fig.~2).  Column densities are given
  in \cm }
\label{tbl_2}
\begin{tabular}{lccccc}
\hline
\noalign{\smallskip}
  &\multicolumn{2}{c}{subsystem $A$} 
  &\multicolumn{2}{c}{subsystem $B$} 
  &\multicolumn{1}{c}{subsystem $C^d$}\\
  &\multicolumn{2}{c}{$-90$ \kms $< v < 0$ \kms} 
  &\multicolumn{2}{c}{0 \kms $< v < 80$ \kms} 
  &\multicolumn{1}{c}{$-20$ \kms $< v < 20$ \kms}\\
Parameter$^a$ & HM96 & S1 & HM96 & S1 &    \\
(1) & (2) & (3) & (4) & (5) & (6)  \\
\noalign{\smallskip}
\hline
\noalign{\smallskip}
$U_0$&8.1E--3&4.5E--3&6.2E--3&4.1E--3& \\
$N_{\rm H}$  & 1.2E18 & 7.7E17 & 7.6E18 & 4.7E18 &  \\
$\sigma_{\rm v}$, \kms & 15.0 & 13.0 & 19.8 & 23.5 &  \\
$\sigma_{\rm y}$& 0.55 & 0.60 & 0.68 & 0.64 &  \\
$Z_{\rm C}$&8.3E--5 & 1.0E--4 & 1.2E--5 & 1.7E--5 &  \\
$Z_{\rm O}$&2.5E--4 & 3.6E--4 & 4.2E--5 & 6.0E--5 &  \\
$Z_{\rm Si}$&1.1E--5 & 1.6E--5 & 1.8E--6 & 2.9E--6 &  \\
$[Z_{\rm C}]$&$-0.47$ &$-0.37$ & $-1.30$ & $-1.15$ & \\
$[Z_{\rm O}]$&$-0.27$ & $-0.10$ & $-1.04$ &$-0.90$ & \\
$[Z_{\rm Si}]$&$-0.46$ & $-0.31$ & $-1.25$ & $-1.04$ &  \\
$N$(H\,{\sc i})&1.4E15 & $(1.4\pm0.4)$E15 & 1.0E16 
& $(1.2\pm0.3)$E16 & $<$3.0E14 \\
$N$(C\,{\sc ii})&3.6E12$^b$ &4.3E12$^b$ & 4.8E12 
& $(7.0\pm2.0)$E12 & $\ldots$ \\
$N$(O\,{\sc ii})&8.1E12$^b$ &1.0E13$^b$ &1.2E13$^b$ 
&2.2E13$^b$ & $\ldots$ \\ 
$N$(Si\,{\sc ii})&4.9E11$^b$ &5.6E11$^b$ & 6.9E11 
& $(9.0\pm3.0)$E11 & $\ldots$ \\
$N$(C\,{\sc iii})&7.0E13 & 5.5E13$^c$ & 7.3E13 & 6.0E13$^c$ 
& $(1.0\pm0.3)$E13$^e$  \\
$N$(O\,{\sc iii})&2.2E14 &2.0E14$^c$ & 2.5E14 & 2.1E14$^c$ & $\ldots$  \\
$N$(Si\,{\sc iii})&5.8E12 & $(6.1\pm0.9)$E12& 8.4E12 & 
$(8.1\pm1.2)$E12 & $\ldots$ \\
$N$(C\,{\sc iv}) &1.7E13&$(1.50\pm0.15)$E13&1.1E13 & 
$(1.1\pm0.1)$E13 & $(1.9\pm0.2)$E13 \\
$N$(O\,{\sc iv}) &5.6E13 &$(6.2\pm2.0)$E13 &4.9E13$^b$ 
& 5.0E13$^b$ & $(2.9\pm0.9)$E14$^e$ \\
$N$(Si\,{\sc iv}) &3.4E12 &$(2.6\pm0.4)$E12 & 2.2E12 & 
$(2.0\pm0.4)$E12 & $\ldots$ \\
$N$(N\,{\sc v})&$\ldots$ &$\ldots$ & $\ldots$ & $\ldots$ &
 $\leq$2.0E12  \\
$N$(O\,{\sc vi}) &$\ldots$ &$\ldots$ &$\ldots$ & $\ldots$ 
 & $(1.10\pm0.25)$E14  \\
$\langle T \rangle$, K & 1.3E4 & 1.5E4 & 2.0E4 & 2.0E4 & \\
$n_{\rm H}$, \cmm & 3.3E--3 & 6.6E--3 & 5.0E--3 & 7.6E--3 & \\ 
$L$, kpc & 0.13 & 0.04 & 0.5 & 0.2 & \\ \noalign{\smallskip} \hline
\noalign{\smallskip} 
\multicolumn{6}{l}{$^aZ_{\rm X} = N_{\rm
    X}/N_{\rm H}$; $[Z_{\rm X}] = \log (N_{\rm X}/N_{\rm H}) - \log
  (N_{\rm X}/N_{\rm H})_\odot$; $^b$calculated using the velocity and
  density}\\ 
\multicolumn{6}{l}{distributions estimated from the metal
  profiles marked by the horizontal bold lines in
  Fig.~1;}\\ 
\multicolumn{6}{l}{$^c$uncertainty cannot be estimated
  since line is corrupted by background
  subtraction;}\\ 
\multicolumn{6}{l}{ $^d$column densities for
    putative \ion{O}{vi} absorption at $-70$ \kms $< v < -20$ \kms
  and at }\\ 
\multicolumn{6}{l}{ $30$ \kms $< v < 50$ \kms are not
    given because of probable blending with some Ly$\alpha$ forest
    lines} \\ 
\multicolumn{6}{l}{ (observed intensities of
    \ion{O}{vi}$\lambda1032$ and $\lambda1037$ are inconsistent);\,
    $^e$can be blended  }
\end{tabular}
\end{table*}

\section{Absorption systems towards \object{HS 0747+4259}}

In the spectrum of \object{HS 0747+4259} we identified over 50
absorption systems, among them 25 with metal lines (Table~1).

From these systems, only a few turned out to be suitable for analysis
by MCI.  The selection criteria included the presence of both lines in
the \ion{C}{iv} and \ion{O}{vi} doublets, and at least one additional
carbon line (e.g. \ion{C}{iii}) in order to fix the ionization
parameter. As an initial guess for the UV background, the HM96 spectra at
each corresponding redshift was used.  The mean intensities of the
ionizing background at $\lambda$ = 912 \AA, $J_{912}$, needed to
calculate the number density of photons and the line-of-sight size of
the absorbing gas, were also taken from HM96.

Note that these $J_{912}$ values only slightly differ (within 20\%)
from those obtained from the proximity effect at $z = 1-2$ (Scott et
al. 2002) and from measurements of the mean radiation flux transmitted
through the Ly$\alpha$ forest (Tytler et al. 2004; Jena et al. 2005).

All computations below were performed with laboratory wavelengths and
oscillator strengths taken from Morton (2003) for $\lambda >$ 912
\AA\, and from Verner et al. (1994) for $\lambda <$ 912 \AA. Solar
abundances were taken from Asplund et al. (2004).
 
The uncertainties of the physical parameter estimates
($U_0$, $N_{\rm H}$, $\sigma_{\rm v}$, $\sigma_{\rm y}$, and $Z_{\rm
  a}$) are about 20\%.

\begin{figure*}[t]
\vspace{0.0cm}
\hspace{0.0cm}\psfig{figure=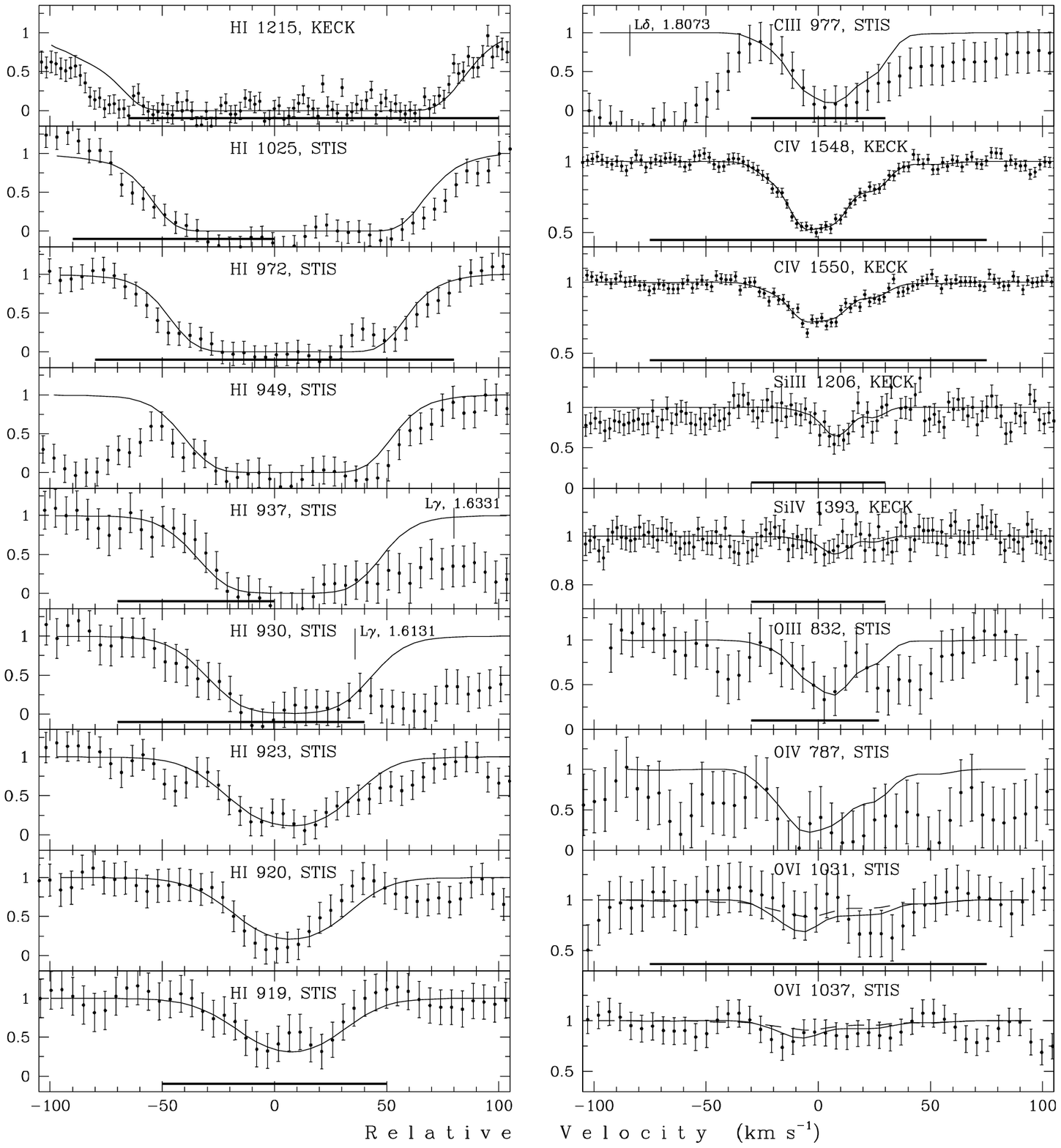,height=18.0cm,width=19.0cm}
\vspace{-0.7cm}
\caption[]{ Same as Fig.~1 but for the \zabs = 1.7301 system.  The
  zero radial velocity is fixed at $z = 1.7301$.  The physical
  parameters corresponding to the background S1 are listed in Table~3,
  Col.3, whereas those obtained with HM96 are in Table~3, Col.2.  The
  central positions of blends are indicated by tick marks.  }
\label{fig3}
\end{figure*}

\subsection{Absorption system at $z = 1.8073$}

This complex absorption system containing lines of different ionic
transitions is shown in Fig.~1.  Unfortunately, some of these lines
are corrupted by bad background subtraction and/or high noise
(\ion{O}{iii} 832, \ion{O}{iv} 788, probably also \ion{C}{iii} 977).
Our first computational runs revealed that the system consists of three
subsystems with different physical parameters: low ionization
subsystems at $-90$ \kms $< v < 0$ \kms (subsystem $A$) and 0 \kms $<
v <$ 80 \kms (subsystem $B$)  seen in \ion{C}{ii}-\ion{C}{iv},
  \ion{Si}{ii}-\ion{Si}{iv} and \ion{O}{ii}-\ion{O}{iv}  and
overlapping high ionization subsystem(s) seen in \ion{C}{iv} and
\ion{O}{vi} at $-20$ \kms $< v < 20$ \kms (subsystem $C$) and perhaps
in \ion{O}{iv}, \ion{O}{vi} at 30 \kms $< v <$ 50 \kms and in
\ion{O}{vi} at  $-70$ \kms $< v <$ $-20$ \kms.

\begin{figure*}[t]
\vspace{0.0cm}
\hspace{0.0cm}\psfig{figure=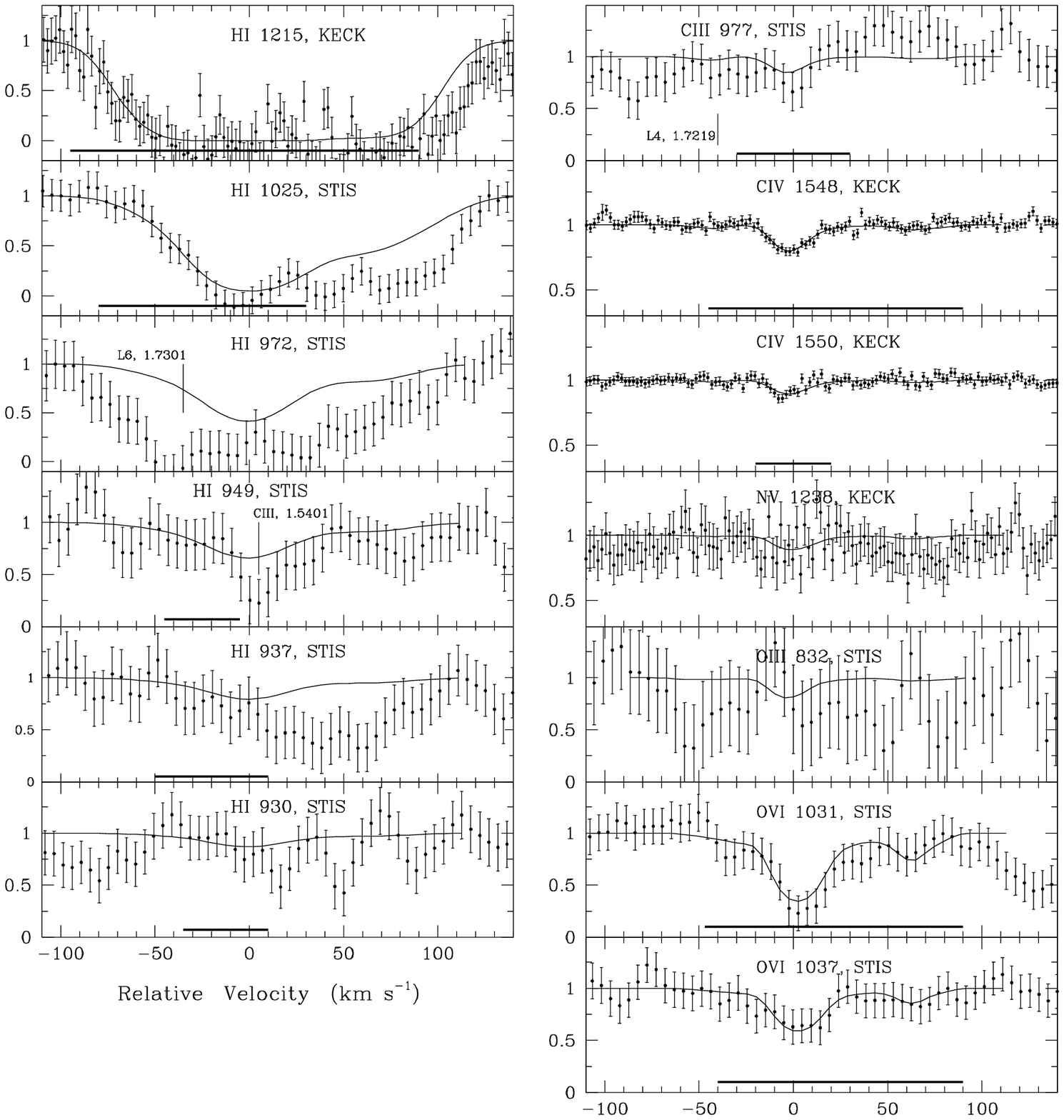,height=18.0cm,width=19.0cm}
\vspace{-1.0cm}
\caption[]{ Same as Fig.~1 but for the \zabs = 1.6131 system.  The
  zero radial velocity is fixed at $z = 1.6131$.  The physical
  parameters corresponding to the background S1 are listed in Table~3,
  Col.5, whereas those obtained with HM96 are in Table~3, Col.4.  The
  central positions of blends are indicated by tick marks. }
\label{fig4}
\end{figure*}

In the following the subsystems $A, B$ and $C$ are treated
separately. Calculations of subsystems $A$ and $B$ were carried out
with the ionizing spectrum of HM96 at $z = 1.8$ (solid line in
Fig.~2).  The obtained physical parameters are given in Table~2, Col.2
and 4.  Due to low S/N it was possible to reproduce almost all line
profiles to within the noise level, i.e. the spectrum of HM96 may be
considered as consistent with the ionization state of both
subsystems. However, some overestimation of \ion{Si}{iv} 1393 in the
subsystem $A$ and underestimation of \ion{C}{ii} 1334 and \ion{Si}{ii}
1260 in the subsystem $B$ hint to probable inadequacy of the adopted
spectral shape (Fig.~1, dashed lines).  Low S/N and bad background
subtraction do not allow us to restore the SED of ionizing radiation
via the directed search procedure described in Agafonova et
al. (2005). However, the spectral shape can be probed by the `trial
and error' method.  For example, the spectrum labeled as S1 in Fig.~2
delivers a better fit to the observed line intensities compared to
HM96. The synthetic line profiles for the subsystems $A$ and $B$
calculated with physical parameters obtained with spectrum S1 (Col.3
and 5 in Table~2) are shown by the smooth lines in Fig.~2. Part of the
\ion{C}{iv} absorption at $-30$ \kms $< v < 15$ \kms was not included
in the calculations because of possible input from the subsystem $C$.

As for the highly ionized subsystem $C$, its hydrogen lines are hidden
in the \ion{H}{i} absorption lines of subsystem $B$.  The upper limit
for the \ion{H}{i} column density along with the column densities for
the \ion{C}{iii} 977, \ion{C}{iv} 1548, 1550, \ion{O}{iv} 832 and
\ion{O}{vi} 1031, 1037 lines identified in this subsystem (intensities
at the expected positions of \ion{C}{iii} and \ion{O}{iv} may be
caused by unidentified blends) are given in Col.6 of Table~2.  The
FWHM of the \ion{O}{vi} lines is 24 \kms which corresponds to $T_{\rm
  kin} \sim 1.6\times10^5$~K if the line broadening is entirely
thermal. According to Sutherland \& Dopita (1993), maximal output of
the \ion{O}{vi} ion in the case of collisional ionization is reached at
$T_{\rm kin} = 3\times10^5$~K and decreases quickly at lower
temperatures.  Thus, this subsystem may be produced either by gas that
has been (shock)-heated and is now cooling and recombining
(non-equilibrial ionization) or by low-density gas in photoionization
equilibrium with an ionizing background hard enough to account for the
observed column density of \ion{O}{vi} (e.g. such as HM96 or S1).

\begin{table*}[t]
\centering
\caption{ Physical parameters of the \zabs = 1.7301 and 1.6131 metal
  absorbers towards \object{HS 0747+4259} derived by the MCI procedure
  with the Haardt \& Madau and modified UV background spectra (marked,
  respectively, by HM96 and S1 in Fig.~2).  Column densities are given
  in \cm }
\label{tbl_3}
\begin{tabular}{lcccc}
\hline
\noalign{\smallskip}
  &\multicolumn{2}{c}{\zabs=1.7301}& \multicolumn{2}{c}{\zabs=1.6131}\\
Parameter$^a$ & HM96 & S1 & HM96 & S1 \\
(1) & (2) & (3) & (4) & (5) \\
\noalign{\smallskip}
\hline
\noalign{\smallskip}
$U_0$& 1.0E--1 & 5.7E--2 & $\ga$2.8E--1 & $\ga$1.3E--1 \\
$N_{\rm H}$&1.9E20 & 9.5E19 & 5.7E19 & 2.5E19 \\
$\sigma_{\rm v}$, \kms & 24.9 & 24.5 & 26.4 & 24.5 \\
$\sigma_{\rm y}$& 0.72 & 0.68 & 0.53 & 0.64 \\
$Z_{\rm C}$&8.1E--7 & 1.6E--6 & 3.1E--6 & 6.1E--6 \\
$Z_{\rm N}$&$\ldots$ & $\ldots$ & $<$2.0E--6 & $<$4.8E--6 \\
$Z_{\rm O}$&2.8E--6 & 5.6E--6 & 1.8E--5 & 3.3E--5 \\
$Z_{\rm Si}$&1.4E--7 & 2.7E--7 & $\ldots$ & $\ldots$ \\
$[Z_{\rm C}]$&$-2.50$ &$-2.18$ & $-1.90$ & $-1.60$ \\
$[Z_{\rm N}]$&$\ldots$ & $\ldots$ & $<-1.5$ &$<-1.1$ \\
$[Z_{\rm O}]$&$-2.24$ & $-1.94$ & $-1.41$ &$-1.17$ \\
$[Z_{\rm Si}]$&$-2.37$ & $-2.11$ & $\ldots$ & $\ldots$ \\
$N$(H\,{\sc i})&1.9E16 & $(1.9\pm0.4)$E16 & 7.3E14 & 
$(7.2\pm2.0)$E14 \\
$N$(C\,{\sc iii}) &4.0E13 & $(3.9\pm0.8)$E13 &  3.6E12 & 
$(3.5\pm1.5)$E12 \\
$N$(O\,{\sc iii}) &1.3E14 & $(1.3\pm0.4)$E14 & 3.9E13$^b$ & 3.0E13$^b$ \\
$N$(Si\,{\sc iii}) &1.7E12&$(1.7\pm0.2)$E12&$\ldots$&$\ldots$ \\
$N$(C\,{\sc iv}) &3.1E13&$(3.10\pm0.15)$E13&1.2E13 & $(1.1\pm0.1)$E13 \\
$N$(O\,{\sc iv}) &2.3E14$^b$ &2.4E14$^b$ &$\ldots$ & $\ldots$ \\
$N$(Si\,{\sc iv}) &9.0E11 &$(8.3\pm3.0)$E11 & $\ldots$ & $\ldots$ \\
$N$(N\,{\sc v}) &$\ldots$ &$\ldots$ & $<$1.6E13 & $<$1.6E13 \\
$N$(O\,{\sc vi}) &2.8E13 &$(4.0\pm1.0)$E13 & 1.3E14 & $(1.30\pm0.25)$E14 \\
$\langle T \rangle$, K & 4.0E4 & 3.9E4 & $\ga$4.7E4 & $\ga$4.5E4 \\
$n_{\rm H}$, \cmm & 3.2E--4 & 5.6E--4 & $\la$1.0E--4 & $\la$2.4E--4 \\ 
$L$, kpc & 200 & 56 & $\ga$190 & $\ga$35 \\
\noalign{\smallskip}
\hline
\noalign{\smallskip}
\multicolumn{5}{l}{$^aZ_{\rm X} = N_{\rm X}/N_{\rm H}$; 
$[Z_{\rm X}] = \log (N_{\rm X}/N_{\rm H}) -
\log (N_{\rm X}/N_{\rm H})_\odot$;
$^b$calculated using the}\\
\multicolumn{5}{l}{corresponding velocity and density distributions 
estimated from the metal profiles}\\ 
\multicolumn{5}{l}{marked by the horizontal bold lines in Figs.~3 and 4}
\end{tabular}
\end{table*}

In general, the parameters obtained for the whole \zabs = 1.8073
absorption complex~-- high metallicity and low ionization compact
clouds ($L \la 0.2$ kpc) embedded in hot highly ionized gas~--
resemble gas observed in a bursted out superbubble by Heckman et
  al. (2001).  The suggested vicinity to a galaxy makes it
interesting to test an ionizing background with significant input from 
stellar radiation, i.e. a spectrum which is softer at $E > 3$ Ryd
compared to the QSO/AGN-dominated spectrum of HM96.  The calculations
show that any type of softer spectra can unambiguously be ruled out,
since they poorly reproduce the observed line intensities and,
moreover, lead to a relative overabundance of carbon compared to
silicon, ([C/Si]$ > 0$). As far as we know, such an overabundance has
never been measured in an \ion{H}{ii} region and is not expected
theoretically.

\begin{figure*}[t]
\vspace{0.0cm}
\hspace{0.0cm}\psfig{figure=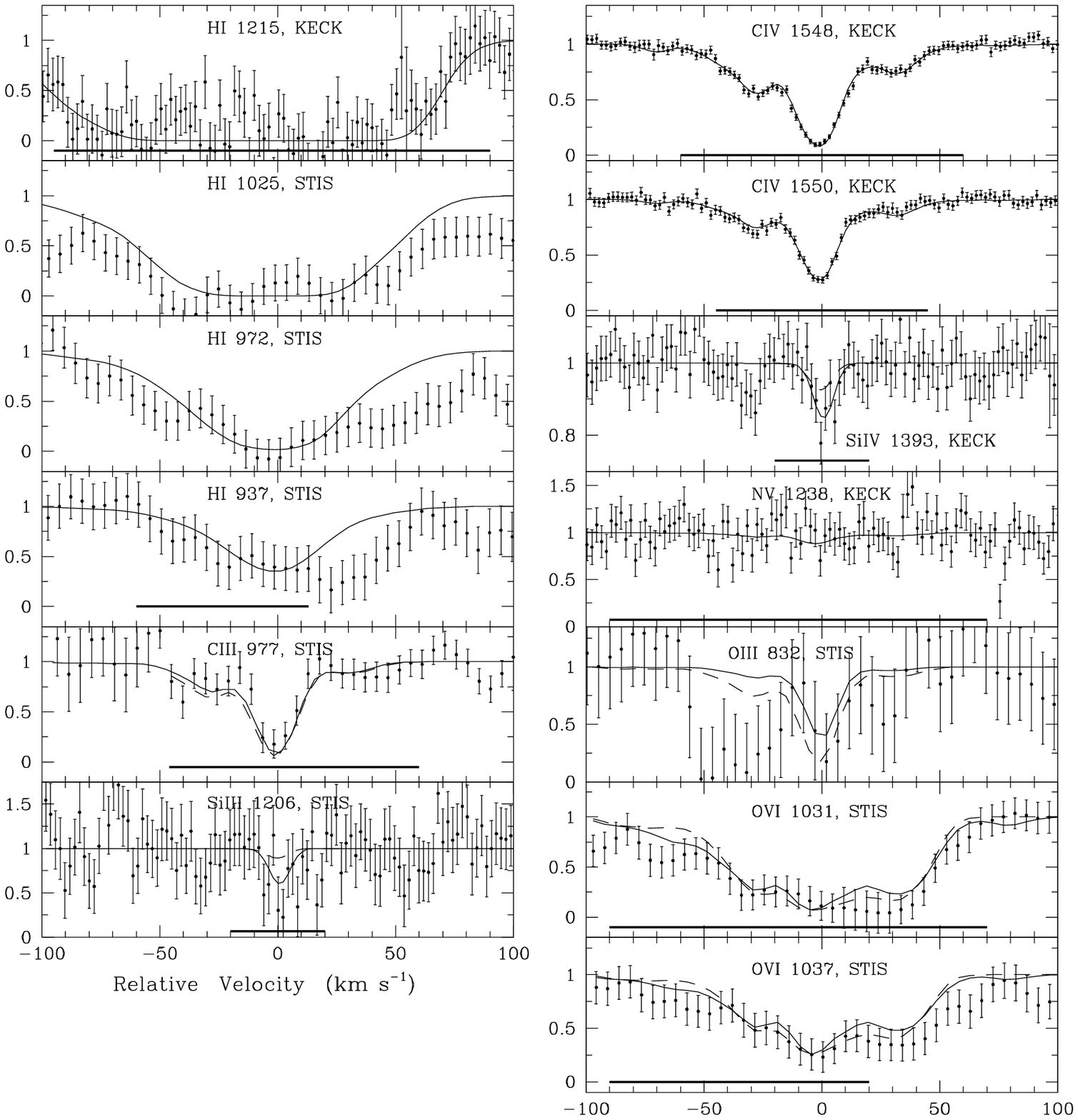,height=18.0cm,width=19.0cm}
\vspace{-1.0cm}
\caption[]{ Same as Fig.~1 but for the \zabs = 1.5955 system.  Smooth
  curves are the synthetic spectra calculated with the ionizing
  background S2 (Fig.~2).  The zero radial velocity is fixed at $z =
  1.5955$.  The physical parameters corresponding to the background S2
  are listed in Table~4, Col.3, whereas those obtained with HM96 are
  in Table~4, Col 2. }
\label{fig5}
\end{figure*}

\subsection{Absorption system at \zabs = 1.7301}

The system exhibits metal lines both in low (\ion{Si}{iii} 1206,
\ion{O}{iii} 832) and high (\ion{C}{iv} 1548, 1550; \ion{O}{vi} 1032,
1037) ionization stages (Fig.~3).  The parameters obtained from the
MCI calculations with the HM96 spectrum are given in Col.2 of Table~3.
With this ionizing background, the observed ratio of \ion{O}{iii} 832
and \ion{O}{vi} 1032, 1037 lines is reproduced only marginally: a good
fit to the \ion{O}{iii} line is accompanied by the underestimation of
\ion{O}{vi} (Fig.~3, dashed lines).  Unfortunately, low S/N at the
positions of these lines does not allow us to unambiguously reject the
adopted spectrum.  However, there is another inconsistency -- the
extremely large linear size of the absorber, $L \sim 200$ kpc.  Linear
sizes of this order of magnitude (hundreds of kpc) are expected for
the filamentary large-scale structure, but the gas in the \zabs =
1.7301 system is too dense (number density $3\times10^{-4}$ \cmm,
overdensity $\delta_{\rm H} \sim 80$) to be attributed to a filament.
According to both observations (Chen et al. 1998, 2001) and
theoretical predictions (Dav\'e et al. 1999), strong Ly$\alpha$
absorbers [$N$(\ion{H}{i}) $> 10^{15}$ \cm] arise at impact parameters
to a galaxy of $r < 50 h^{-1}$ kpc.  Thus, the linear size of the
\zabs = 1.7301 absorber may be similar.

The ionizing spectrum of HM96 was modified in a way to enhance the
\ion{O}{vi}/\ion{O}{iii} ratio and to increase the gas number density
    [i.e. to decrease the mean ionization parameter $U_0$, see
      eq.(1)].  The fractional abundance of \ion{O}{iii} decreases
    with the enhancement of the intensity at $3 < E < 4$ Ryd and the
    fraction of \ion{O}{vi} becomes larger with higher intensity at $E
    > 4$ Ryd.  One of the acceptable spectral shapes is shown in
    Fig.~2 by the short-dashed line labeled as S1.  Results obtained
    with S1 are given in Col.3 of Table~3, and the synthetic profiles
    are plotted in Fig.~3 by the smooth lines.  Note that with the
    same normalizing intensity $J_{912}$ the spectrum S1 gives nearly
    four times smaller linear size than the spectrum of HM96.

It should be stressed that the shape S1 is only one of the possible
solutions for the \zabs = 1.7301 system since the low S/N of most of
the spectral data allows us to vary the intensity at $E > 3$ Ryd over
a broad range.  Thus, the exact spectral shape cannot be recovered.
Nevertheless, we can conclude that the absorption lines observed in
this system favor spectra with increased intensity at $E > 3$ Ryd as
compared to the HM96 model.

\begin{figure*}[t]
\vspace{0.0cm}
\hspace{0.0cm}\psfig{figure=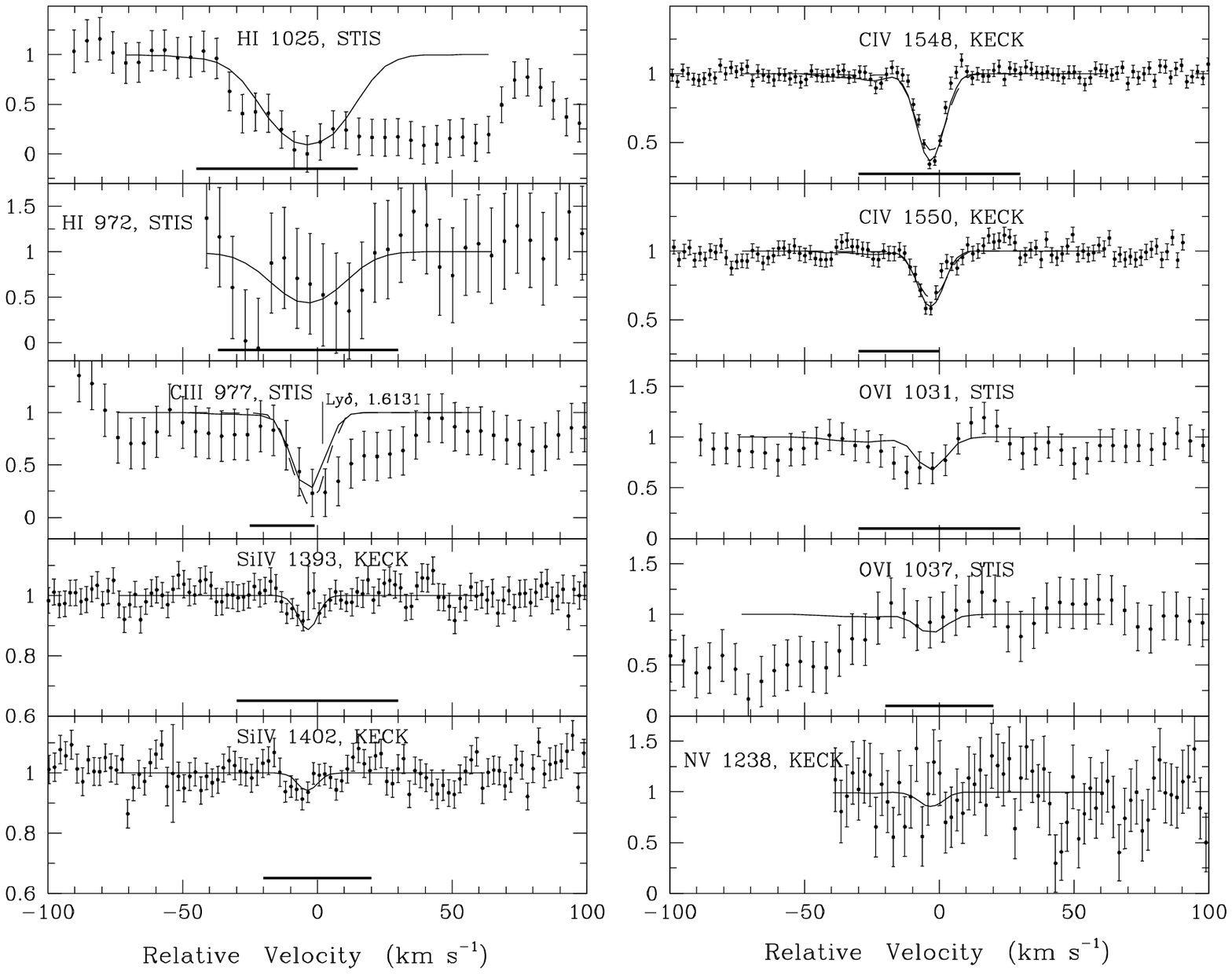,height=18.0cm,width=19.0cm}
\vspace{-5.0cm}
\caption[]{ Same as Fig.~1 but for the \zabs = 1.5401 system.  The
  zero radial velocity is fixed at $z = 1.5401$.  The smooth curves
  are the synthetic spectra calculated with the ionizing background S2
  (Fig.~2).  The physical parameters corresponding to the background
  S2 are listed in Table~4, Col.5, whereas those obtained with HM96
  are in Table~4, Col.4.  The central positions of blends are
  indicated by tick marks. }
\label{fig6}
\end{figure*}

\subsection{Absorption system at \zabs = 1.6131}

This highly ionized system exhibits several hydrogen lines along with
a weak \ion{C}{iv} doublet and strong lines of \ion{O}{vi} 1031, 1037
(Fig.~4). At the expected position of the \ion{Si}{iv} 1393 line a
clear continuum is present.  The observed intensity at the position of
the \ion{C}{iii} 977 line can be used only as an upper limit on the
\ion{C}{iii} absorption since this line lies in a noisy part of the
spectrum and is partly contaminated by Ly$\delta$ from the \zabs =
1.7219 system.  This upper limit, along with a safe upper limit on the
abundance ratio [O/C]$ < 0.5$ known from measurements in Galactic and
extragalactic \ion{H}{ii} regions and in metal poor halo stars (Henry
et al. 2000; Akerman et al. 2004), are utilized to estimate the
physical parameters in the absorber. The parameters obtained from the
MCI calculations with the HM96 spectrum are listed in Col.4 of
Table~3, whereas those obtained with the spectrum S1 (Fig.~2) are
given in Col.5.  The synthetic line profiles are identical in both
cases and are shown by the smooth lines in Fig.~4.

The spectrum of HM96 leads to a linear size of hundreds of kpc and,
hence, places the absorber in some filamentary structure.  The
spectrum S1 is harder at $E > 3$ Ryd and provides a line-of-sight
size several times smaller, which allows us to attribute the absorber to
a galactic halo.

\begin{figure*}[t]
\vspace{0.0cm}
\hspace{0.0cm}\psfig{figure=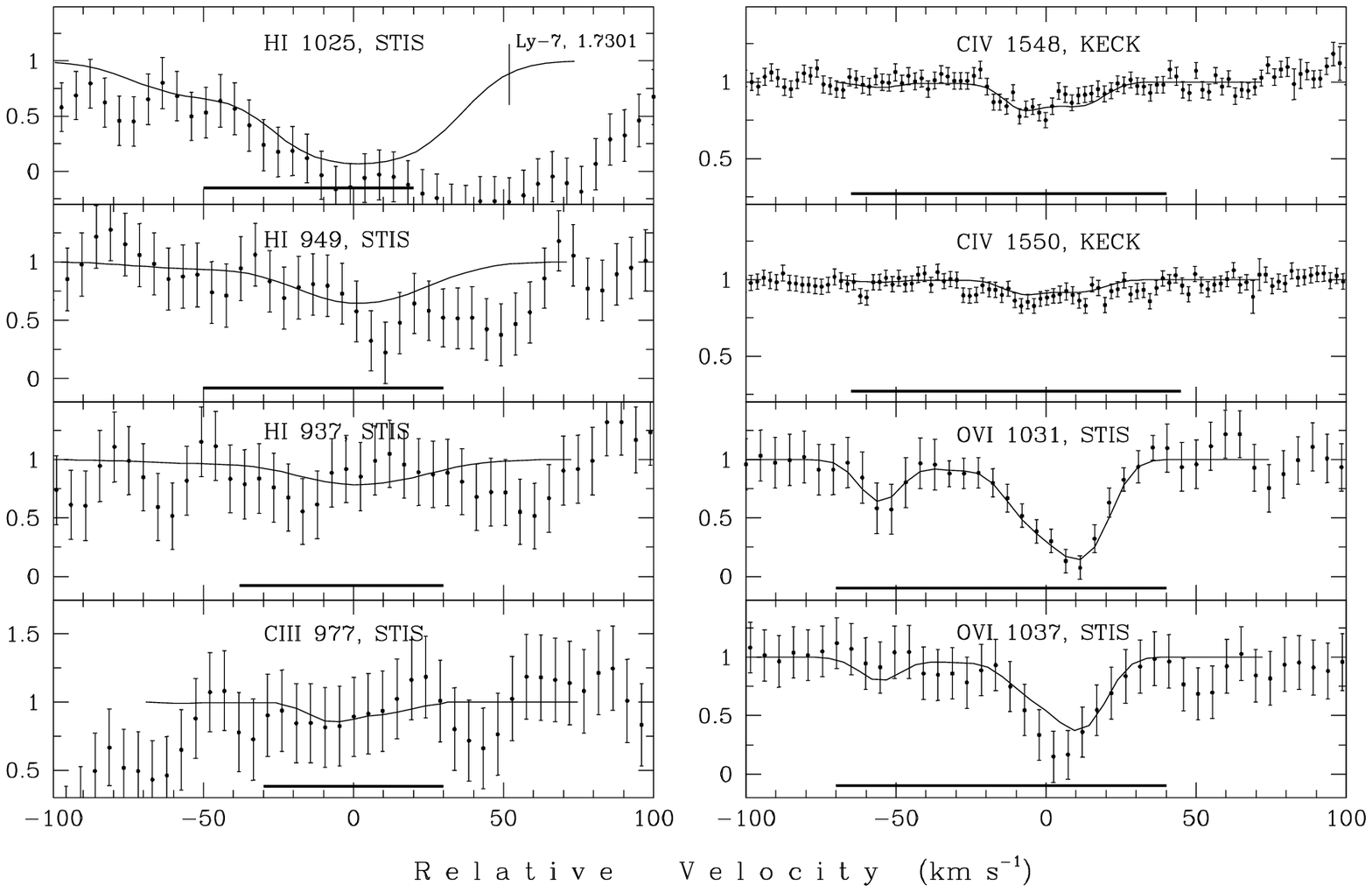,height=18.0cm,width=19.0cm}
\vspace{-7.5cm}
\caption[]{ Same as Fig.~1, but for the \zabs = 1.4649 system.  The
  zero radial velocity is fixed at $z = 1.4649$.  The physical
  parameters corresponding to the backgrounds HM96, S1, and S2 are
  listed in Table~4, Cols.6,7, and 8, respectively.  The central
  positions of blends are indicated by tick marks. }
\label{fig7}
\end{figure*}

Both results are obtained assuming photoionization equilibrium for all
observed ions.  However, collisional ionization of \ion{O}{vi} should
be tested as well.  From the apparent FWHMs of the available lines: 
\ion{H}{i} (60 \kms), \ion{C}{iv} (25 \kms), and \ion{O}{vi} (35 \kms),
and assuming pure thermal line broadening, we obtain kinetic
temperatures of $T_{\rm HI} = 7.6\times10^4$~K, $T_{\rm CIV} =
1.4\times10^5$~K, and $T_{\rm OVI} = 3.8\times10^5$~K.  Such diverse
temperatures cannot occur in the same gas and thus turbulent
broadening must be significant.  The possibility of  collisionally
ionized \ion{O}{vi} (with corresponding broad and shallow hydrogen
lines hidden in the observed profile of \ion{H}{i}) surrounds cooler
gas producing \ion{H}{i} and \ion{C}{iv} absorption seems to be quite
unprobable since the lines of \ion{C}{iv} and \ion{O}{vi} are well
aligned in velocity space and have similar profiles.  The case of
non-equilibrium ionization can be ruled out as well since the
recombination time of \ion{O}{vi} is much shorter than that of
hydrogen: for example, at $T_{\rm kin} \sim 3\times10^5$~K it is about
60 times faster (Osterbrock 1974).  Then the bulk of \ion{O}{vi} should
come to photoionization equilibrium prior to \ion{H}{i} and we would
observe a weak and shallow \ion{H}{i} Ly$\alpha$ along with strong
\ion{O}{vi} lines.  This contradicts the observed profiles in the
\zabs = 1.6131 system.  Therefore photoionization equilibrium is the
most probable assumption.

It should be noted that \ion{O}{vi} lines trace more rarefied gas
not seen in \ion{C}{iv} and, hence, can be broader.  On the other
hand, the \ion{H}{i} profile can be wider than \ion{O}{vi} since some
low density regions produce feeble \ion{O}{vi} absorption not
distinguishable from the noise. Thus, we can estimate the lower limit
on the turbulent velocity dispersion from the comparison of the
\ion{H}{i} and \ion{O}{vi} lines: $\sigma_{\rm turb} > 19$ \kms.

Taking into account such high turbulent velocities, the placement of
the \zabs = 1.6101 absorber in a galactic halo, i.e. within boundaries
where it remains gravitationally bound to a galaxy, seems to be more
reasonable than its origin in the filamentary structure with no
significant sources of the line broadening apart from the Hubble
expansion.  Thus, ionizing spectra that are harder at $E > 4$ Ryd as
compared to that of HM96 are preferable for this absorber.

A few words should be said about the high upper limit on the nitrogen
abundance since measurements usually show [N/C]$ < -0.5$ for metal
poor gas (Centuri\'on et al. 2003).  Unfortunately, the \ion{N}{v}
1238 line in the absorber under study lies in a very noisy part of the
Keck spectrum with many broad continuum undulations of unidentified
nature.  It is possible that this low contrast feature is not
completely due to \ion{N}{v} 1238 absorption.

\begin{table*}[t]
\centering
\caption{ Physical parameters of the \zabs = 1.5955, 1.5401 and 1.4649
  metal absorbers towards \object{HS 0747+4259} derived by the MCI
  procedure with the Haardt \& Madau and modified UV background
  spectra (marked, respectively, by HM96, S1, and S2 in Fig.~2).
  Column densities are given in \cm }
\label{tbl_4}
\begin{tabular}{lccccccc}
\hline
\noalign{\smallskip}
  &\multicolumn{2}{c}{\zabs=1.5955}& \multicolumn{2}{c}{\zabs=1.5401} &
\multicolumn{3}{c}{\zabs=1.4649}\\
Parameter$^a$ & HM96 & S2 & HM96 & S2 & HM96 & S1  & S2 \\
(1) & (2) & (3) & (4) & (5) & (6) & (7) & (8) \\
\noalign{\smallskip}
\hline
\noalign{\smallskip}
$U_0$&2.4E--1&5.5E--2&1.8E--1&2.7E--2&$\ga$3.4E--1&
$\ga$1.8E--1&$\ga$8.8E--2 \\
$N_{\rm H}$ & 1.0E20 & 1.6E19 & 2.2E18 & 4.7E17 
& 6.8E19 & 2.6E19 & 1.3E19 \\
$\sigma_{\rm v}$, \kms & 32.7 & 29.3 & 13.1 & 12.2 
 & 22.0 & 21.0 & 23.2 \\
$\sigma_{\rm y}$& 0.64 & 0.60 & 0.83 & 0.41 
 & 0.46 & 0.60 & 0.35 \\
$Z_{\rm C}$&6.5E--6 & 4.4E--5 & 2.6E--5 & 1.5E--4 
 & 3.4E--6 & 6.8E--6 & 2.0E--5 \\
$Z_{\rm N}$&$<$4.5E--7 & $<$3.0E--6 & $<$1.1E--5 & $<$3.6E--5 
 & $\ldots$ & $\ldots$ & $\ldots$ \\
$Z_{\rm O}$&3.2E--5 & 1.5E--4 & 7.4E--5 & 2.7E--4 
  & 1.8E--5 & 3.7E--5 & 6.0E--5 \\
$Z_{\rm Si}$&2.2E--6 & 7.6E--6 & 3.5E--6 & 1.8E--5 
 & $\ldots$ & $\ldots$ & $\ldots$ \\
$[Z_{\rm C}]$&$-1.57$ &$-0.74$ & $-0.98$ & $-0.21$ 
 & $-1.86$ & $-1.55$ & $-1.10$ \\
$[Z_{\rm N}]$&$<-2.1$ & $<-1.3$ & $<-0.74$ &$<-0.2$ 
& $\ldots$ & $\ldots$ & $\ldots$ \\
$[Z_{\rm O}]$&$-1.15$ & $-0.50$ & $-0.80$ &$-0.23$ 
 & $-1.40$ & $-1.10$ & $-0.90$ \\
$[Z_{\rm Si}]$&$-1.17$ & $-0.63$ & $-0.99$ &
$-0.26$ & $\ldots$ & $\ldots$ & $\ldots$ \\
$N$(H\,{\sc i})&2.7E15 & $(2.5\pm0.5)$E15 & 3.5E14 & $(3.4\pm0.6)$E14 
 & 6.8E14 & 6.8E14 & $(7.0\pm2.0)$E14 \\
$N$(C\,{\sc iii}) &3.8E13 & $(3.6\pm0.6)$E13&1.6E13$^b$ & 1.3E13$^b$ 
 & 2.6E12 & 2.6E12 &  $(1.8\pm0.9)$E12 \\
$N$(O\,{\sc iii}) &1.9E14$^b$ & 9.0E13$^b$ & $\ldots$ & $\ldots$ &
$\ldots$ & $\ldots$ & $\ldots$ \\  
$N$(Si\,{\sc iii}) & 2.7E11 & 1.1E12&$\ldots$&$\ldots$ 
 & $\ldots$ & $\ldots$  & $\ldots$ \\
$N$(C\,{\sc iv}) &9.0E13&$(9.1\pm0.4)$E13&1.45E13 & $(1.75\pm0.15)$E13 
 & 1.2E13 & 1.1E13 & $(1.2\pm0.1)$E13 \\
$N$(Si\,{\sc iv}) &5.2E11 &$(1.00\pm0.15)$E12 & 6.1E11 & $(5.9\pm0.6)$E11 
 & $\ldots$ & $\ldots$ & $\ldots$ \\
$N$(N\,{\sc v}) &$<$7.5E12 &$<$9.5E12 & $<$4.0E12 & $<$4.0E12 &
 $\ldots$ & $\ldots$ & $\ldots$ \\
$N$(O\,{\sc vi}) &4.5E14 &$(4.2\pm0.8)$E14 & 1.7E13 & $(1.9\pm0.4)$E13 
 & 1.7E14 & 1.7E14 & $(1.70\pm0.35)$E14 \\
$\langle T \rangle$, K & 3.5E4 & 3.0E4 & 2.8E4 & 2.0E4 
 & $\ga$4.8E4 & $\ga$4.8E4 & $\ga$3.3E4 \\
$n_{\rm H}$, \cmm & 1.2E--4 & 5.1E--4 & 1.9E--4 & 1.2E--3 
& $\la$6.7E--5 & $\la$1.5E--4 & $\la$3.0E--4\\ 
$L$, kpc & 280 & 10 & 4 & 0.13 & $\ga$280 & $\ga$60 & $\ga$15  \\
\noalign{\smallskip}
\hline
\noalign{\smallskip}
\multicolumn{8}{l}{$^aZ_{\rm X} = N_{\rm X}/N_{\rm H}$; 
$[Z_{\rm X}] = \log (N_{\rm X}/N_{\rm H}) -
\log (N_{\rm X}/N_{\rm H})_\odot$;
$^b$calculated using the velocity and density
distributions }\\
\multicolumn{8}{l}{
estimated from the metal profiles 
marked by the horizontal bold lines in Fig.~5 (\zabs=1.59)
and in Fig.~6 (\zabs=1.54)}\\ 
\end{tabular}
\end{table*}

\subsection{Absorption system at \zabs = 1.5955}

Absorption lines identified in this system are shown in Fig.~5. Note
the very strong lines of \ion{O}{vi} 1032, 1037 observed together with
lines of \ion{Si}{iii} 1206 and \ion{Si}{iv} 1393 (\ion{Si}{iv} 1402
is blended).  The physical parameters derived by the MCI assuming the
HM96 ionizing background are listed in Col.2 of Table~4, and the
corresponding synthetic profiles are shown by dashed lines in Fig.~5.
This ionizing spectrum is not optimal: for the distribution of
ionization parameters determined by the \ion{C}{iii}, \ion{C}{iv} and
\ion{O}{vi} lines it significantly underproduces intensities of
silicon lines (or, equivalently, delivers the ratio Si/O significantly
higher than solar, which is inconsistent with measurements in
\ion{H}{ii} regions) and it gives too large of a linear size. For this
system non-equilibrium ionization can also be ruled out: the
broadening of the \ion{C}{iii} 977, \ion{C}{iv} 1548, 1550,
\ion{O}{vi} 1031, 1037 and hydrogen lines is similar.  This points to
significant contributions from turbulent velocities and, hence, to
temperatures of about $10^4$~K (see previous subsection).  Moreover,
there is a safe upper limit on the nitrogen abundance [N/C]$ < -0.5$
obtained from the \ion{N}{v} 1238 line: since \ion{N}{v} traces the
same gas as \ion{O}{vi}, any `overionization' of this gas would result
in an artificial enhancement of the nitrogen abundance calculated
under the assumption of equilibrium ionization. Thus, we look for a
more appropriate ionizing background.

Taking into account information obtained with the HM96 spectrum and
using the technique described in Reimers et al. (2005) and Agafonova
et al. (2005), we searched for a background that would ensure a higher
ratio of fractional ionizations $\Upsilon_{\rm SiIV}/\Upsilon_{\rm
  CIV}$ at the ionization parameter $U$ determined from the
\ion{C}{iii}, \ion{C}{iv} and \ion{O}{vi} absorption.  Additionally,
this $U$ should be low enough to produce a physically reasonable
linear size for the absorber under study. Other constraints to the
solution were the inequalities [Si,O/C]$ < 0.5$.

The resulting spectrum is shown in Fig.2 by the dashed line and marked
as S2.  It is significantly harder in the whole range $E > 1$ Ryd
compared to both the initial spectrum of HM96 and the spectrum S1 used
for the absorbers described in the preceding subsections.  The
physical parameters obtained from the MCI calculations with the S2
spectrum are listed in Col.3 of Table~4, and the corresponding
synthetic profiles are plotted by the smooth lines in Fig.~5.  Note
that the derived element abundance ratios [Si,O,N/C] are consistent
with measurements in Galactic and extragalactic \ion{H}{ii} regions
(e.g., Matteucci 2003).

The assortment and column densities of lines observed in the \zabs =
1.5955 system resemble those identified in some highly ionized
high-velocity clouds (HVCs) at $z = 0$ (Collins et al. 2004). Perhaps
we are observing in the \zabs = 1.5955 absorber a high-redshift analog of
the local HVCs.

One important difference between the spectra S2 and S1 should be
emphasized. The shape S1 represents only one possible solution from a
wide range of acceptable shapes since the low S/N for lines observed
in the \zabs = 1.8073, 1.7301 and 1.6131 systems does not permit us to
recover the SED with reasonable accuracy.  Conversely, lines observed
in the \zabs = 1.5955 system can be self-consistently described only
with a quite narrow range of possible ionizing spectra and, hence, the
spectrum S2 is determined much more accurately.  In this context it is
interesting to test the shape S2 on the absorption systems at \zabs =
1.8073 and 1.7301 exhibiting many metal lines. The result is
negative~-- this spectral shape is inconsistent with line strengths
observed in these systems.  Thus, the ionizing radiation at $z < 2$
seems to undergo strong spectral variations.

\subsection{Absorption system at \zabs = 1.5401}

This absorption system is seen in \ion{H}{i} Ly$\beta$ and Ly$\gamma$
and in metal lines shown in Fig.~6 (Ly$\alpha$ and \ion{Si}{iii} 1206
fall in the wavelength gap between the HST and Keck spectra).  Note
the unusually narrow \ion{C}{iv} lines with FWHM = 12.5 \kms.

The parameters derived from the MCI calculations with the HM96
ionizing spectrum are listed in Col.4 of Table~4.  The synthetic
profiles are shown in Fig.~6 by dashed lines.  This spectrum is not
completely consistent with the observed line intensities: it
overestimates \ion{C}{iii} (the intensity of \ion{C}{iii} should be lower
than the apparent intensity due to blending with Ly$\delta$ from the
\zabs = 1.6131 system) and significantly underestimates the profiles
of \ion{C}{iv} 1548, 1550 in the center.

The spectra S1 and S2 were tried as well. S1 was rejected for the same
reason as the spectrum of HM96~-- poor reproduction of carbon lines.
The spectrum S2 turned out to be consistent with the data.  The
physical parameters obtained with this background are listed in Col.5
of Table~4, and the corresponding synthetic line profiles are plotted
in Fig.~6 (smooth lines).

This absorption system has characteristics which are quite unusual for
intergalactic absorbers: metallicity of 0.6 solar and a sub-kiloparsec
linear size. Probably it was ejected from a nearby galaxy located
transverse to the line of sight.  Note that the hard ionizing
continuum derived for both \zabs = 1.5955 and \zabs = 1.5401 absorbers
(separated by 6500 \kms) may be considered as a hint to the presence
of an AGN in the vicinity of these systems (cf. S2 and the AGN-type
spectrum of Mathews \& Ferland in Fig.~2).

\subsection{Absorption system at \zabs = 1.4649}

This system is almost identical to that at \zabs = 1.6131 (the lines
\ion{N}{v} 1238, 1242 fall in the gap between Keck and STIS spectra).
The physical parameters obtained from calculations with the HM96, S1
and S2 spectra are listed in Col.6, 7 and 8 of Table~4,
respectively. The synthetic profiles coincide for all three cases and
are shown by the smooth lines in Fig.~7.  The system exhibits a
similar high turbulent velocity as found at \zabs = 1.6131.  Therefore
its origin in a filamentary structure is less favorable than an origin
in a galactic halo and, hence, the harder spectra S1 and S2 are
preferable to that of HM96.

\subsection{Frequency distribution of \ion{O}{vi} absorbers}

The detection limit of \ion{O}{vi}$\lambda1302$ lines in the
spectrum of \object{HS 0747+4259} corresponds to $N$(\ion{O}{vi}) $
\ga 2\times10^{13}$ \cm\, ($W_{\rm rest} \ga 30$ m\AA).  In addition
to the 6 absorbers described above, a further 10 metal systems in the
redshift range 1.07~-1.87 exhibit \ion{O}{vi} $\lambda1031$ absorption
with the second line $\lambda1037$ blended with Ly$\alpha$ forest
absorption.  However, the identification of \ion{O}{vi} in these
systems is certain due to the simultaneous detection of the
\ion{C}{iv} and \ion{O}{iv} $\lambda787$ lines (in absorbers with
\zabs $> 1.7$).  Table~5 lists the redshifts and column densities for
the detected \ion{O}{iv} lines.

\begin{table}[t]
\centering
\caption{\ion{O}{vi} column densities 
}
\label{tbl_5}
\begin{tabular}{lccc}
\hline
\noalign{\smallskip}
  &\zabs& $N$(\ion{O}{vi}), \cm & Comments\\
\noalign{\smallskip}
\hline
1& 1.8617 & $(3.2\pm0.8)$E13 & $\lambda1037$ blended \\
2& 1.8533 & $(3.0\pm0.8)$E13 & $\lambda1037$ blended \\
3& 1.8073 & $(1.1\pm0.3)$E14 & \\
4& 1.7790 & $(2.5\pm0.8)$E13 & $\lambda1037$ below detection limit \\
5& 1.7744 & $(4.0\pm1.0)$E13 & $\lambda1037$ blended \\
6& 1.7302 & $(4.0\pm1.0)$E13 & \\
7& 1.7154 & $(2.5\pm0.8)$E13 & $\lambda1037$ blended \\
8& 1.6331 & $(7.0\pm1.5)$E13 & $\lambda1037$ blended \\
9& 1.6134 & $(1.3\pm0.3)$E14 & \\ 
10& 1.5955& $(4.2\pm0.8)$E14 & \\
11& 1.5401& $(1.9\pm0.4)$E13 & \\
12& 1.4856& $(1.1\pm0.3)$E14 & $\lambda1037$ blended \\
13& 1.4648& $(1.7\pm0.4)$E14 & \\
14& 1.2910& $(5.0\pm3.0)$E13 & noisy spectrum, S/N $\sim 3$ \\
15& 1.1896& $(7.0\pm3.0)$E13 & $\lambda1037$ blended \\
16& 1.0778& $(5.0\pm3.0)$E13 & noisy spectrum, S/N $\sim 3$ \\
\noalign{\smallskip}
\hline
\end{tabular}
\end{table}

With the absorption distance of $\Delta X = 1.27$\, $(q_0 =
1/2)$\footnote{$\Delta X = \frac{2}{3}\left[(1+z_1)^{3/2} -
    (1+z_2)^{3/2}\right]$}, the number of encountered \ion{O}{vi}
absorbers translates into an absorber frequency of $\Delta{\cal
  N}_{\rm OVI}/\Delta X = 13$.  This is almost two times higher than
$\Delta{\cal N}_{\rm OVI}/\Delta X = 7$ measured for the range $1.21
<$ \zabs $< 1.67$ $(\Delta X = 0.72)$ in the spectrum of \object{HE
  0515-4414} (Reimers et al. 2001).  Thus, the \ion{O}{vi} absorber
statistics at \zabs $\sim 1.5$ is subject to strong variations from
sightline to sightline.  The mass density of the \ion{O}{vi}
absorbers, $\sum N$(\ion{O}{vi})$/\Delta X$, shows even stronger
variations: $(1.2\pm0.3)\times10^{15}$ \cm\, along the present
sightline compared to $3.0\times10^{14}$ \cm\, towards \object{HE
  0515--4414}.  Note, however, that the bulk of the integrated column
densities in \object{HS 0747+4259} comes from the strong \ion{O}{vi}
absorbers clustered at $1.46 <$ \zabs $< 1.63$ where hard ionizing
background was recovered.  For comparison, at low redshift, $z \la
0.15$, the \ion{O}{vi} absorber frequency ($W_{\rm res} > 30$ m\AA)
averaged over 31 FUSE sightlines is\, $\Delta{\cal N}_{\rm OVI}/\Delta
X = 17\pm3$\, (Danforth \& Shull 2005).  We recognize, however, such a
comparison neglects the fact that the \ion{O}{vi} systems at $z \sim
1.5$ are mainly photoionized, while those in the local universe may
originate in the collisionally ionized intergalactic medium.

\section{Conclusions}

In this work, we investigate the \ion{O}{vi} absorption line systems
detected in the HST/STIS and Keck/HIRES spectra of the QSO \object{HS
  0747+4259}.  Six systems \ion{O}{vi} with \zabs = 1.46-1.81 reveal a
sufficient number of lines to perform a quantitative analysis of the
shape of the UV background ionizing radiation.  The results obtained
lead to the following conclusions.

\begin{enumerate}
\item[1.] Spectral energy distributions with a significantly ($>
  2.5$ times) higher intensity at $E > 4$ Ryd as compared to the model
  spectrum by HM96 are preferable for the absorption systems
  studied. Similar conclusion have been obtained in Agafonova et
  al. (2005) for the absorber at \zabs = 1.9426 towards \object{J
    2233--606}.  Such SEDs are suggested by the deficiency of strong
  Ly$\alpha$ absorbers [$N$(\ion{H}{i}) $> 10^{15}$ \cm] at $z <
  2$. On the other hand, the metagalactic SEDs obtained at $z \sim 3$
  (Agafonova et al. 2005) show an intensity at $E > 4$ Ryd consistent
  with that of HM96. This confirms the conclusion of Kim et al. (2001)
  that absorbers with strong Ly$\alpha$ lines evolve faster than weak
  ones.
\item[2.]The spectral shapes recovered for the absorbers at \zabs =
  1.8073, 1.7301 and at \zabs = 1.5955, 1.5401 are different with the
  latter being significantly harder in the whole range $E > 1$ Ryd.
  There are theoretical predictions (Dav\'e et al. 1999) that in
  general the evolution of the Ly$\alpha$ absorbers flattens at $z <
  1.7$.  However, this flattening would not lead to the increased
  ionizing flux.  Thus, the observed variation of the spectral shape
  is probably of local nature and may be caused by the presence of a
  QSO/AGN close the line of sight (cf. Jakobsen et al. 2003).
\item[3.]The spectral shapes at $z < 2$ do not show signs of a
  significant contribution from galactic (stellar) radiation.
  According to Bianchi et al. (2001), an escape fraction $f_{\rm esc}
  = 0.05$ results in a galactic contribution to the UV background at
  912 \AA\ equal to that of QSOs.  Such galactic input would
  noticeably soften the ionizing spectrum in the range $E >$ 3 Ryd
  (see, e.g., Giroux \& Shull 1997).  Taking into account the hardness
  of the recovered spectra at $E > 3$ Ryd , the escape fraction of
  galactic photons at $z < 2$ should be well below 0.05. This is in
  line with $f_{\rm esc} < 0.04$ estimated by Fernandez-Soto et
  al. (2003) from measurements of 27 galaxies with $1.9 < z < 3.5$.
\item[4.] In the systems considered, the main ionization
  mechanism for all observed ions including \ion{O}{vi} is
  photoionization by the background UV radiation.  The contribution of
  turbulence to the line broadening is significant and the
  temperatures estimated from the widths (FWHMs) of the \ion{O}{vi}
  lines are well below the range where collisional ionization is
  dominant.  High \ion{O}{vi} column densities ($N > 10^{14}$ \cm) can
  originate in non-equilibrium gas which has been shock-heated and is
  now cooling and recombining (Heckman et al. 2002).  Such conditions
  can be valid for very particular absorption system at \zabs = 1.8073
  (subsystem $C$), but no other systems show signs of being
  non-equilibrial. Similar conclusions about the ionization of high
  redshift \ion{O}{vi} are reported by Carswell et al. (2002),
  Bergeron et al. (2002), and Levshakov et al. (2003).
\end{enumerate}

\begin{acknowledgements}
We thank staff of the W.M. Keck observatory, including instrument
specialist Randy Cambell and observing assistants Gabrelle Saurage and
Gary Puniwai.  The work of I.I.A. and S.A.L. is supported by the RFBR
grant 03-02-17522 and the RLSS grant 1115.2003.2.  C.F. is supported
by the Verbundforschung of the BMBF/DLR grant No.~50 OR 9911 1.
D.T. and D.K. were supported in part by NASA StScI grant GO-9040.02
and by NSF grant AST-0098731.

\end{acknowledgements}

\end{document}